\documentstyle[epsf]{elsart}

\newcommand{\be}{\begin{equation}}
\newcommand{\ee}{\end{equation}}
\newcommand{\ba}{\begin{array}}
\newcommand{\ea}{\end{array}}
\newcommand{\bea}{\begin{eqnarray}}
\newcommand{\eea}{\end{eqnarray}}
\newcommand{\bean}{\begin{eqnarray*}}
\newcommand{\eean}{\end{eqnarray*}}
\newcommand{\lp}{\left(}
\newcommand{\rp}{\right)}
\newcommand{\ls}{\left[}
\newcommand{\rs}{\right]}
\newcommand{\lc}{\left\{}
\newcommand{\rc}{\right\}}
\renewcommand{\d}{\mbox{\rm d}}

\newcommand{\mPi}{{\mit\Pi}}

\newcommand{\refp}[1]{(\ref{#1})}
\newcommand{\ds}{\displaystyle}
\newcommand{\bmv}[1]{\mbox{\boldmath $#1$}}

\newcommand{\tensor}[1]{\stackrel{\leftrightarrow}{#1}}

\begin{document}
\begin{frontmatter}
\title{\bf Normal solution to the Enskog-Landau kinetic equation.
	Boundary conditions method}

\author{A.E.Kobryn, I.P.Omelyan, M.V.Tokarchuk}

\address{Institute for Condensed Matter Physics\\
	Ukrainian National Academy of Sciences,\\
	1~Svientsitskii St., UA--290011 Lviv--11, Ukraine}

\begin{abstract}
Nonstationary and nonequilibrium processes are considered on the basis of an
Enskog-Landau kinetic equation using a boundary conditions method. A
nonstationary solution of this equation is found in the pair collision
approximation. This solution takes into account explicitly the influence of
long-range interactions. New terms to the transport coefficients are
identified. An application of the boundary conditions method to
hydrodynamic description of fast processes is discussed.
\end{abstract}

\end{frontmatter}

The development of methods to construct a theory for nonequilibrium
processes in dense gases, liquids and plasmas is an important direction in
the modern theoretical physics. Moreover, the construction of kinetic
equations for such classical and quantum systems still remains to be a
major problem in the kinetic theory. It is complicated additionally in
the case of dense gases, liquids and plasmas, where kinetics and
hydrodynamics are closely connected and should be considered simultaneously
\cite{c1,c2,c3,c4,c5}.

An approach for construction of kinetic equations from the first principles
of statistical mechanics, namely from the Liouville equation, has been
developed in \cite{c6,c7}. Another approach for obtaining kinetic equations
for{\it dense} systems, which is
based on ideas of papers \cite{c6,c7}, has also been proposed \cite{c1} and
generalized in \cite{c2,c3}. Here, the formulation of modified boundary
conditions for the BBGKY hierarchy is used taking into account
corrections connected with local conservation laws. On the basis of this
sequential approach, a new Enskog-Landau kinetic equation has been
obtained for an one-component system of charged hard spheres.
There is a considerable interest of an application of this kinetic equation
for description of transport processes in dense systems of charged particles
as well as in ion melts and electrolyte solutions. The normal solution and
transport coefficients for this equation have been found in the paper
\cite{c8} using the Chapman-Enskog method. The same approach has been
used for a many-component system of charged hard spheres in the paper
\cite{c9} with more detailed calculations for a two-component system as
well. At the same time, as is well known, the Chapman-Enskog method allows
one to find the transport coefficients in a stationary case only. Similar
drawbacks are peculiar to the Grad method \cite{c10,c11} which is oftenly
used to solve kinetic equations next to the Chapman-Enskog method.

In this paper, the Enskog-Landau kinetic equation for a system of char\-ged
hard spheres is investigated. To find the normal solution in a
nonstationary case, the so-called boundary conditions method is used,
which has been introduced in \cite{c12,c13}. As a result,
transport coefficients equations in the nonstationary case are written.
A limiting case of the stationary process is considered. A brief comparison
of the obtained transport coefficients with those known previously
from the Chapman-Enskog method is given.

Let us consider the Enskog-Landau kinetic equation for a one-component
sys\-tem of charged hard spheres \cite{c8}:
%
%
\be
\left\{\frac\partial{\partial t}+\bmv{v}_1
	\frac\partial{\partial r_1}\right\}
	f_1\left(x_1;t\right)=I_E\left(x_1;t\right)+
	I_{MF}\left(x_1;t\right)+I_L\left(x_1;t\right),
	\label{e1}
\ee
where $f_1(x_1;t)$ is the one-particle distribution function.
The right-hand side of this equation is the so-called generalized
Enskog-Lan\-dau collision integral, where each term can be considered as a
separate collision integral. Their structure are as follows:

$I_E\left( x_1;t\right)$ is the collision integral of the Enskog theory RET
\cite{c14}:
%
%
\bea
\lefteqn{\ds
I_E\left(x_1;t\right)=\sigma^2\int\d\hat{\bmv{r}}_{12}\;\d\bmv{v}_2\;
	\Theta\left(\hat{\bmv{r}}_{12}\bmv{g}\right)
	\left(\hat{\bmv{r}}_{12}\bmv{g}\right)\times{}}
	\label{e2}\\
&&\Big\{g_2\left(\bmv{r}_1,\bmv{r}_1+\hat{\bmv{r}}_{12}\sigma;t\right)
	f_1\left(\bmv{r}_1,\bmv{v}_1^{\prime};t\right)f_1\left(\bmv{r}_1+
	\hat{\bmv{r}}_{12}\sigma,\bmv{v}_2^{\prime};t\right)-{}
	\nonumber\\
&&g_2\left(\bmv{r}_1,\bmv{r}_1-\hat{\bmv{r}}_{12}\sigma;t\right)
	f_1\left(\bmv{r}_1,\bmv{v}_1;t\right)
	f_1\left(\bmv{r}_1-\hat{\bmv{r}}_{12}\sigma,
	\bmv{v}_2;t\right)\Big\},\nonumber
\eea
where $\sigma$ is a hard sphere diameter, $\bmv{g}$ denotes a vector of
relative velocity for two particles, $\hat{\bmv{r}}_{12}$ is a unit vector
along the direction between centres of particles 1 and 2,
\bean
\lefteqn{
\ba{ll}
\ds\bmv{v}_1'=\bmv{v}_1+\hat{\bmv{r}}_{12}
	(\hat{\bmv{r}}_{12}\cdot\bmv{g}),
&\qquad\bmv{g}=\bmv{v}_2-\bmv{v}_1,\\
\bmv{v}_2'=\bmv{v}_2-\hat{\bmv{r}}_{12}
	(\hat{\bmv{r}}_{12}\cdot\bmv{g}),
&\qquad\hat{\bmv{r}}_{12}=|\bmv{r}_{12}|^{-1}\bmv{r}_{12};\\
\ea}
\eean

$I_{MF}\left( x_1;t\right)$ is the collision integral of the kinetic mean
field theory KMFT \cite{c15,c16}:
%
%
\bea
\lefteqn{\ds I_{MF}\left(x_1;t\right)=\frac 1m\int\d x_2\;
	\frac{\partial{\mit\Phi}^l
	\left(|\bmv{r}_{12}|\right)}{\partial\bmv{r}_1}
	\frac\partial{\partial\bmv{v}_1}
	g_2\left(\bmv{r}_1,\bmv{r}_2;t\right)
	f_1\left(x_1;t\right)f_1\left(x_2;t\right),}\label{e3}
\eea
where ${\mit\Phi}^l\lp|\bmv{r}_{12}|\rp$ is a long-range part of the
interparticle interaction potential;

$I_L\left( x_1;t\right)$ is generalized Landau collision integral
\cite{c2,c8}:
%
%
\bea
\lefteqn{\ds I_L\left(x_1;t\right)=\frac 1{m^2}
	\frac{\partial}{\partial\bmv{v}_1}
	\int\d x_2\;
	g_2\left(\bmv{r}_1,\bmv{r}_2;t\right)\left[
	\frac{\partial{\mit\Phi}^l\left(|\bmv{r}_{12}|\right)}
	{\partial\bmv{r}_{12}}\right]\times{}}
	\label{e4}\\
&&\left[\int\limits_{-\infty}^0\d t'\;
	\frac{\partial{\mit\Phi}^l\left(|\bmv{r}_{12}+
	\bmv{g}t^{\prime }|\right) }{\partial\bmv{r}_{12}}\right]
	\left\{\frac\partial{\partial\bmv{v}_1}-
	\frac\partial{\partial\bmv{v}_2}\right\}f_1\left(x_1;t\right)
	f_1\left(x_2;t\right).\nonumber
\eea
It is necessary to note that the quasiequilibrium binary correlation function
$g_2$ takes into account the full interaction potential (hard core part
plus long-range Coulomb tail).

One of a major problem at the correct derivation and solution of kinetic
equations is their consistency with local conservation laws of particle
density (or mass), momentum, total energy and substantiation of
hydrodynamic equations and incomprehensible calculation of transport
coefficients via molecular parameters. These conservation laws for classical
systems in general have the structure as in \cite{c17}.

To find a solution of the Enskog-Landau kinetic equation (\ref{e1})
using one or another method, there is necessary to take the advantage of
local conservation laws in corresponding approximations. So doing the
expressions for kinetic coefficients will be defined through
densities for momentum flow tensor $\tensor{\mPi}(\bmv{r};t)$ and energy
flow vector $\bmv{j}_{\cal E}(\bmv{r};t)$ on the basis of solution
$f_1(x_1;t)$ and corresponding approximations for
$g_2(\bmv{r}_1,\bmv{r}_2;t)$. As far as we find the solution that
corresponds to linear hydrodynamical transport processes by gradients of
thermodynamical parameters, densities of
momentum flow tensor $\tensor{\mPi}(\bmv{r};t)$ and energy flow vector
$\bmv{j}_{\cal E}(\bmv{r};t)$ could be determined immediately with the help
of kinetic equation \refp{e1} without general formulas from \cite{c17}.
To this end it is convenient similarly to \cite{c2}, to introduce the
following hydrodynamical parameters: density $n(\bmv{r}_1;t)$ (or mass
density $\rho(\bmv{r}_1;t)$), hydrodynamical velocity $\bmv{V}(\bmv{r}_1;t)$
and density of kinetic energy $\omega_k(\bmv{r}_1;t)$.
Multiplying initial kinetic equation (\ref{e1}) by hydrodynamical parameters
and integrating with respect to $\bmv{v}_1$, one can obtain the
equations for these parameters in the form:
%
%
\bea
\lefteqn{\ds\frac 1{\rho\left(\bmv{r}_1;t\right)}
	\frac{\d}{\d t}\rho\left(\bmv{r}_1;t\right)}
	\qquad\qquad\qquad\qquad
	&=&-\frac\partial{\partial\bmv{r}_1}\bmv{V}\left(\bmv{r}_1;t\right),
	\label{e6}\\
	&& \nonumber\\
\lefteqn{\ds\rho\left(\bmv{r}_1;t\right)
	\frac{\d}{\d t}\bmv{V}\left(\bmv{r}_1;t\right)}
	\qquad\qquad\qquad\qquad
	&=&-\frac\partial{\partial\bmv{r}_1}:\tensor{P}
	\left(\bmv{r}_1;t\right),
	\label{e7} \\
\lefteqn{\ds\rho\left(\bmv{r}_1;t\right)
	\frac{\d}{\d t}w_k\left(\bmv{r}_1;t\right)}
	\qquad\qquad\qquad\qquad
	&=&-\frac\partial{\partial\bmv{r}_1}
	\bmv{q}\left(\bmv{r}_1;t\right)-
	\tensor{P}\left(\bmv{r}_1;t\right):
	\frac\partial{\partial\bmv{r}_1}\bmv{V}\left(\bmv{r}_1;t\right),
	\label{e8}
\eea
%
%
where
\bea
\lefteqn{\ds\tensor{P}\left(\bmv{r}_1;t\right)=
	\tensor{P}^k\left(\bmv{r}_1;t\right)+
	\tensor{P}^{hs}\left(\bmv{r}_1;t\right)+
	\tensor{P}^{mf}\left(\bmv{r}_1;t\right)+
	\tensor{P}^l\left(\bmv{r}_1;t\right),}
	\label{e9}\\
\lefteqn{\ds\bmv{q}\left(\bmv{r}_1;t\right)=
	\bmv{q}^k\left(\bmv{r}_1;t\right)+
	\bmv{q}^{hs}\left(\bmv{r}_1;t\right)+
	\bmv{q}^{mf}\left(\bmv{r}_1;t\right)+
	\bmv{q}^l\left(\bmv{r}_1;t\right)}
	\nonumber
\eea
are the total stress tensor and vector of heat flow correspondingly.
They have additive structure and contain several terms, each of them is
stipulated by the influence from one of collision integrals \cite{c2,c8}:
$\tensor{P}^{hs}$ and $\bmv{q}^{hs}$
by Enskog collision integral (\ref{e2}),
$\tensor{P}^{mf}$ and $\bmv{q}^{mf}$
by collision integral of KMFT (\ref{e3}),
$\tensor{P}^l$ and $\bmv{q}^{l}$
by Landau collision integral (\ref{e4}),
$\tensor{P}^k$ and $\bmv{q}^{k}$
are pure kinetic contributions only.
$\tensor{P}^l\lp\bmv{r}_1;t\rp$ and $\bmv{q}^{l}\lp\bmv{r}_1;t\rp$
are new terms in the structure of \refp{e9} in comparison with
results of \cite{c18}:
%
%
\bea
\lefteqn{\ds\tensor{P}^l\left(\bmv{r}_1;t\right)={}}\label{e14}\\
&&\frac{Z^4e^4}{m}\int\d\bmv{v}_1\;
	\bmv{v}_1\frac \partial{\partial\bmv{v}_1}\int\d x_2\;
	\frac{\bmv{r}_{12}\bmv{r}_{12}}{r_{12}^5}\cdot
	\frac{\bmv{r}_{12}}{g}
	\lc\frac{\partial}{\partial\bmv{v}_1}-
	\frac{\partial}{\partial\bmv{v}_2}\rc
	\int\limits_0^1\d\lambda\;F^l,\nonumber
\eea
%
%
%
%
%
\bea
\lefteqn{\ds\bmv{q}^l\left(\bmv{r}_1;t\right)={}}\label{e17}\\
&&\frac{Z^4e^4}{2m}\int\d\bmv{v}_1\;
	c_1^2\frac\partial{\partial\bmv{v}_1}\int\d x_2\;
	\frac{\bmv{r}_{12}\bmv{r}_{12}}{r_{12}^5}\cdot
	\frac{\bmv{r}_{12}}{g}
	\left\{\frac\partial{\partial\bmv{v}_1}-
	\frac\partial{\partial\bmv{v}_2}\right\}
	\int\limits_0^1\d\lambda\;F^l,\nonumber
\eea
$F^l=g_2f_1f_1$ \cite{c8}. A short comment is needed for \refp{e8}. First of
all equation \refp{e8} is a balance equation for a kinetic part of total
energy. To write the conservation law for total energy it is necessary to
know also two-particle distribution function $f_2(x_1,x_2;t)$ next to
one-particle one, because the potential part of total energy is expressed
via $f_2$. The Enskog-Landau kinetic equation in ``pair collision''
approximation has been obtained from the BBGKY hierarchy with a modified
boundary condition in \cite{c2}, where the expression for $f_2$ is also
pointed out. An average value for the potential energy and its flow one
should be calculated on the basis of this expression. Then, adding it to the
balance equation \refp{e8}, one can obtain the conservation law for total
energy.

We shall construct a normal solution to the Enskog-Landau kinetic
equation (\ref{e1}) using the boundary conditions method \cite{c12,c13}.
Following this me\-thod, let us bring into the right-hand side of equation
(\ref{e1}) an infinity small source with $\varepsilon\to +0$:
%
%
\bea
\lefteqn{\ds\left\{\frac\partial{\partial t}+
	\bmv{v}_1\frac\partial{\partial \bmv{r}_1}\right\}
	f_1\left(x_1;t\right)={}}\label{e18}\\
&&\ds I_E\left(x_1;t\right)+I_{MF}\left(x_1;t\right)+
	I_L\left(x_1;t\right)-
	\varepsilon\left(f_1\left(x_1;t\right)-
	f_1^{\left(0\right)}\left(x_1;t\right)\right),
	\nonumber
\eea
where $f_1^{(0)}\lp x_1;t\rp$ is some already known one-particle
distribution function satisfying equations (\ref{e6}) -- (\ref{e8})
for parameters of reduced description of our system. Then the solution
can be found in the form $f_1\lp x_1;t\rp=f_1^{\lp 0\rp}\lp x_1;t\rp+
\delta f\lp x_1;t\rp$ and search of the normal solution implies treatment of
the correction $\delta f\lp x_1;t\rp$. Substituting $f_1(x_1;t)$ into
(\ref{e18}), one can obtain:
%
%
\bea
\lefteqn{\ds\left\{\frac\partial{\partial t}+
	\bmv{v}_1\frac\partial{\partial\bmv{r}_1}+\varepsilon\right\}
	\delta f+\frac D{Dt}f_1^{\left(0\right)}=
	I_{MF}\left(f_1^{\left(0\right)}\right)+
	I_{MF}\left(\delta f\right)+{}}\label{e19}\\
&&I_E\left(f_1^{\left(0\right)},f_1^{\left(0\right)}\right)+
	I_L\left(f_1^{\left(0\right)},f_1^{\left(0\right)}\right)+
	I_E\left(f_1^{\left(0\right)},\delta f\right)+
	I_E\left(\delta f,f_1^{\left(0\right)}\right)+{}
	\nonumber\\
&&I_L\left(f_1^{\left(0\right)},\delta f\right)+
	I_L\left(\delta f,f_1^{\left(0\right)}\right)+
	I_E\left(\delta f,\delta f\right)+
	I_L\left(\delta f,\delta f\right).
	\nonumber
\eea
Conventional signs used in the equation \refp{e19} are obvious
\cite{c2,c8,c9}. Also the fact was taken into account about
$I_{MF}(x_1;t)$, collision integral (\ref{e3}), which is a functional of
one-particle distribution function only. Terms with the subscript $E$ are
nonlocal, therefore in further calculations we should take their
expansion with respect to the local one-particle distribution function and
cut-off this expansion by terms with degrees higher than $\delta f$. In the
case when terms with subscripts $MF$ and $L$ also mean nonlocal functionals,
one should apply mentioned above procedure to them too. Let us combine some
terms in (\ref{e19}):

\begin{flushleft}
\begin{tabular}{@{}lp{6cm}}
\\
\ $\ds I_E\left(\delta f\right)=I_E\left(f_1^{\left(0\right)},\delta f\right)
	+I_E\left(\delta f,f_1^{\left( 0\right)}\right)$ &
linearized nonlocal Enskog collision functional,\\
\ $\ds I_L\left(\delta f\right)=I_L\left(f_1^{\left(0\right)},\delta f\right)
	+I_L\left(\delta f,f_1^{\left(0\right)}\right)$ &
linearized Landau collision func\-ti\-o\-nal.
\ \\
\end{tabular}
\end{flushleft}
\vspace{1ex}
Now let us designate
$L_t\left( \delta f\right) =I_E^{\left( 0\right) }\left( \delta f\right)
+I_{MF}\left( \delta f\right) +I_L\left( \delta f\right)$ and introduce an
operator $S\left( t,t^{\prime }\right)$ with the following properties:
\[
\frac\partial{\partial t}S\left(t,t^{\prime}\right)=
	L_t\left(\delta f\right)S\left(t,t^{\prime}\right),\qquad
	S\left(t,t^{\prime}\right)|_{t^{\prime}=t}=1.
\]
Using these properties of operator $S\lp t,t'\rp$, one can represent equation
(\ref{e19}) in an integral form.
Having correction $\delta f(x;t)$ in an integral form, it is easy to cross to
itemizing procedure for finding it in corresponding approximation. For
example, it can be organized in the following way:
%
%
\bea
\lefteqn{\delta f^{(k+1)}\left(x_1;t\right)=
	\int\limits_{-\infty}^t\d t'\;
	\e^{-\varepsilon(t-t')}S\lp t,t'\rp
	\left\{-\frac D{Dt}f_1^{\lp 0\rp}-
	\bmv{v}_1\frac\partial{\partial\bmv{r}_1}\delta f^{(k)}+{}\right.}
	\label{e22}\\
&&I_E\left(f_1^{\left(0\right)},f_1^{\left(0\right)}\right)+
	I_{MF}\left(f_1^{\left(0\right)}\right)+
	I_L\left(f_1^{\left(0\right)},f_1^{\left(0\right)}\right)+
	I_E^{(1)}(\delta f^{(k)})\Bigg\}_{t'},
	\nonumber
\eea
where subscript $t'$ at the bottom of right brase means that integrated
expression is a function of $t'$. An additional condition to find
$\delta f(x;t)$ is the evident limit
$\ds\lim_{t\to-\infty}\delta f(x;t)=0$.
In order to construct the $(k+1)$-th approximation it is necessary to use
the fact that $\ds\delta f\big|_{k=0}=0$ and the conservation laws
(or equations for reduced description parameters) in $k$-th approximation.
To realize this procedure a zeroth approximation for the one-particle
distribution function $f_1^{(0)}(x_1;t)$ is needed. In the case of spherical
charged particles, $f_1^{(0)}(x_1;t)$ can be chosen as the local-equilibrium
Maxwell distribution function
\[
f_1^{\left(0\right)}\left(x_1;t\right)=n\left(\bmv{r}_1;t\right)
	\left(\frac m{2\pi kT\left(\bmv{r}_1;t\right)}\right)^{3/2}
	\exp\left\{-\frac{mc_1^2\left(\bmv{r}_1;t\right)}
	{2kT\left(\bmv{r}_1;t\right)}\right\}.
\]

Let us find a correction to the distribution function $f_1^{(0)}(x_1;t)$
using itemizing procedure (\ref{e22}). Calculating and obtaining of
conservation laws \refp{e6}, \refp{e7} and equation \refp{e8}, we should
take into account the following relations:
%
%
\be
\ba{@{}l@{}}
\ds g_2\left(\bmv{r}_1,\bmv{r}_2;t\right)\equiv
	g_2\Big(\bmv{r}_1,\bmv{r}_2;n(t),\beta(t)\Big)
	\rightarrow g_2\Big(\bmv{r}_{12};n(t),\beta(t)\Big),
	\phantom{\int_1^1}\\
F\ds\rightarrow F^{\left(0\right)}=
	g_2\Big(\bmv{r}_{12};n(t),\beta(t)\Big)
	f_1^{\left(0\right)}\left(x_1;t\right)
	f_1^{\left(0\right)}\left(\bmv{r}_1,\bmv{v}_2;t\right),\\
\ea\label{e23}
\ee
where $g_2\big(\bmv{r}_{12};n(t),\beta(t)\!\big)$ is the binary
quasiequilibrium correlation function, which depends on relative distance
between particles. We obtain for stress tensor and heat flow vector:
%
%
\bea
\lefteqn{\ds\tensor{P}^k=\tensor{I}P^k,}
	\qquad\qquad\qquad\qquad
	&&P^k=nkT,\label{e24}\\
\lefteqn{\ds\tensor{P}^{hs}=\tensor{I}P^{hs},}
	\qquad\qquad\qquad\qquad
	&&P^{hs}=\frac 23
	\pi n^2\sigma^3kTg_2\left(\sigma|n,\beta\right),
	\label{e25}\\
\lefteqn{\ds\tensor{P}^{mf}=\tensor{I}P^{mf},}
	\qquad\qquad\qquad\qquad
	&&P^{mf}=-\frac 23\pi
	\left( nZe\right) ^2\int\limits_\sigma^\infty
	\frac{\d r}r\;g_2\left(r|n,\beta\right),
	\label{e26}\\
\lefteqn{\ds\tensor{P}^l=0,}
	\qquad\qquad\qquad\qquad
	&&\bmv{q}^k=\bmv{q}^{hs}=\bmv{q}^{mf}=
	\bmv{q}^l=0.\label{e27}
\eea
In these expressions $\tensor{I}$ is the unit tensor, $\tensor{P}^l$ and
$\bmv{q}^l$ are equal to zero because the integration between symmetrical
limits goes over odd function. As far as calculated components
$\tensor{P}^k(\bmv{r}_1;t)$, $\tensor{P}^{hs}(\bmv{r}_1;t)$,
$\tensor{P}^{mf}(\bmv{r}_1;t)$ and $\tensor{P}^l(\bmv{r}_1;t)$
(\ref{e24}) -- (\ref{e27}) are known, one can write total
pressure in the zeroth approximation:
\[
P=nkT\lp 1+\frac 23\pi n\sigma^3g_2(\sigma|n,\beta)\rp-
	\frac 23\pi(nZe)^2\int\limits_\sigma^\infty
	\frac{\d r}{r}\;g_2(r|n,\beta).
\]
Calculating expressions in brackets on the right hand side in
(\ref{e22}), one can write total expression for correction $\delta f(x_1;t)$
in first approximation:
%
%
\bea
\lefteqn{\ds\delta f^{(1)}\left(x_1;t\right)=-\int\limits_{-\infty }^t\d t'\;
	\e^{-\varepsilon(t-t')}S\left(t,t'\right)
	\Bigg[f_1^{(0)}\left(x_1;t\right)\times{}}
	\label{e31}\\
\lefteqn{\ds\left\{\left(1+\frac 25\pi n\sigma^3
	g_2\left(\sigma|n,\beta\right)\right)
	\left[\frac{mc_1^2}{2kT}-\frac 52\right]\bmv{c}_1
	\frac{\partial}{\partial\bmv{r}_1}\ln T\left(\bmv{r}_1;t\right)+{}
	\right.}\nonumber\\
\lefteqn{\ds\left.\left.\left(1+\frac 4{15}\pi n\sigma^3
	g_2\left(\sigma|n,\beta\right)\right)
	\frac m{kT}\left[\bmv{c}_1\bmv{c}_1-\frac 13c_1^2\tensor{I}\right]:
	\frac{\partial}{\partial\bmv{r}_1}
	\bmv{V}\left(\bmv{r}_1;t\right)\right\}\right]_{t'}.}
	\nonumber
\eea
Terms related to short-range interactions only contribute evidently into the
correction in this approximation. Contrary to the kinetic theory of dilute
gases particle sizes take part here \cite{c6,c7,c11}, where
particles are considered as point-like objects. Nevertheless, the influence
of both long-range and short-range parts of interactions are also ``hidden''
in operator $S(t,t')$ (through operator $L_t$). Formally, the expression
\refp{e31} looks completely the same as the correction in \cite{c18}.
But a difference lies in the structure of the operator $S(t,t')$.

Having total expression for correction $\ds\delta f(x_1;t)$ in the first
approximation (\ref{e31}) one can calculate conservation laws \refp{e6},
\refp{e7} and equation \refp{e8} in the same approximation. Therefore,
it is necessary, first, to obtain relations for determining quantities
\refp{e8} in which the correction (\ref{e31}) can be engaged.
For $\tensor{P}^{k\; 1}\lp\bmv{r}_1;t\rp$ we obtain:
%
%
\be
\tensor{P}^{k\; 1}\left(r_1;t\right)=\tensor{I}P^k-\int\limits_{-\infty }^t
	\d t'\;\e^{-\varepsilon(t-t')}M^k\left(t,t'\right)
	\ls\tensor{S}\;\rs_{t'},\label{e32}
\ee
where $S_{\alpha\beta}$ is a velocity shift tensor,
%
%
\bea
\lefteqn{\ds M^k\left(t,t'\right)=\frac m5
	\int\d\bmv{v}_1\;\bmv{c}_1\bmv{c}_1
	S\left(t,t'\right)\times{}}\label{e34}\\
&&\left[f_1^{(0)}\left(x_1;t\right)\left(1+\frac 4{15}\pi n\sigma^3
	g_2\left(\sigma|n,\beta\right)\right)\frac m{kT}
	\left(\bmv{c}_1\bmv{c}_1-
	\frac 13c_1^2\tensor{I}\right)\right]_{t^{\prime }}
	\nonumber
\eea
is a kernel of kinetic part of the transport equations.

For calculating $\tensor{P}^{hs\; 1}\lp\bmv{r}_1;t\rp$,
$\tensor{P}^{mf\; 1}\lp\bmv{r}_1;t\rp$ and
$\tensor{P}^{l\; 1}\lp\bmv{r}_1;t\rp$, we have to expand
$F^{hs}$, $F^{mf}$ and $F^{l}$ on inhomogeneity
and deviation $\delta f\lp x_1;t\rp$ and keep in the series initial
terms only. The expansion for $F^{mf}$, $F^{l}$ reads the same as for
$F^{hs}$ with changing $g_2(\sigma|n,\beta)\to g_2(\bmv{r}_{12}|n,\beta)$,
$\hat{\bmv{r}}_{12}\to |\bmv{r}_{12}|^{-1}\bmv{r}_{12}$,
$\sigma\to|\bmv{r}_{12}|$. The calculations show:
%
%
\bea
\lefteqn{\ds\tensor{P}^{hs\;1}\left(\bmv{r}_1;t\right)=
	\tensor{P}^{hs}-\frac 49n^2\sigma^4
	g_2\left(\sigma|n,\beta\right)\sqrt{\pi mkT}\left[\frac 65\tensor{S}+
	\left(\nabla\cdot\bmv{V}\right)\tensor{I}\right]-{}}
	\label{e36}\\
&&\frac 4{15}\pi n\sigma^3g_2\left(\sigma|n,\beta\right)
	\int\limits_{-\infty}^t\d t'\;\e^{-\varepsilon(t-t')}
	M^k\left(t,t'\right)\ls\tensor{S}\;\rs_{t'},
	\nonumber\\
\lefteqn{\tensor{P}^{mf\;1}\left(\bmv{r}_1;t\right)=\tensor{I}P^{mf}.}
	\label{e37}
\eea
A mean field influence into the total stress tensor remains the same as in
zeroth approximation. Similar situation arises as to
$\tensor{P}^{l\; 1}\left(\bmv{r}_1;t\right)$:
%
%
\be
\tensor{P}^{l\;1}\left(\bmv{r}_1;t\right)=
	\tensor{P}^{l}\left(\bmv{r}_1;t\right)=0.
	\label{e38}
\ee
Total expression for stress tensor in the first approximation is a sum of
(\ref{e32}), (\ref{e36}), (\ref{e37}) and (\ref{e38}):
\bean
\lefteqn{\ds\tensor{P}(\bmv{r}_1;t)=\tensor{I}P(\bmv{r}_1;t)-
	\frac 49n^2\sigma^4
	g_2\lp\sigma|n,\beta\rp\sqrt{\pi mkT}\left[\frac 65\tensor{S}+
	\left(\nabla\cdot\bmv{V}\right)\tensor{I}\right]-{}}\\
&&\lp 1+\frac 4{15}\pi n\sigma^3g_2\left(\sigma|n,\beta\right)\rp
	\int\limits_{-\infty}^t\d t'\;\e^{-\varepsilon(t-t')}
	M^k\left(t,t'\right)\ls\tensor{S}\;\rs_{t'}.\\
\eean

The calculations for heat flow vectors give:
%
%
\bea
\lefteqn{\ds\bmv{q}^{k\; 1}\left(\bmv{r}_1;t\right)=-
	\int\limits_{-\infty }^t\d t'\;
	\e^{-\varepsilon(t-t')}L^k\left(t,t'\right)
	\left[\frac 1T\nabla T\right]_{t'},}
	\label{e39}\\
\lefteqn{\ds\bmv{q}^{hs\;1}\left(\bmv{r}_1;t\right)=-\frac 23n^2\sigma^4g_2
	\left(\sigma|n,\beta\right)\sqrt{\frac{\pi k^3T}m}\nabla T
	\left(\bmv{r}_1;t\right)-{}}\label{e40}\\
&&\frac 25\pi n\sigma^3g_2(\sigma|n,\beta)
	\int\limits_{-\infty }^t\d t'\;\e^{-\varepsilon(t-t')}
	L^k\left(t,t'\right)\left[\frac 1T\nabla T\right]_{t'},
	\nonumber\\
\lefteqn{\ds\bmv{q}^{l\;1}\left(\bmv{r}_1;t\right)=
	\bmv{q}^{l}\left(\bmv{r}_1;t\right)=0.}
	\label{e41}
\eea
Here
%
%
\bea
\lefteqn{\ds L^k\left(t,t'\right)=
	\frac 13\int\d\bmv{v}_1\;c_1\frac{mc_1^2}2
	S\left(t,t'\right)\times{}}
	\label{e42}\\
&&\left[f_1^{(0)}\left(x_1;t\right)\left(1+\frac 25\pi n\sigma^3
	g_2\left(\sigma|n,\beta\right)\right)\left(\frac{mc_1^2}{2kT}-
	\frac 52\right)c_1\right]_{t'}
	\nonumber
\eea
is another kernel of kinetic part of transport equations. Total expression
for heat flux vector is a sum of (\ref{e39}) -- (\ref{e41}):
\bean
\lefteqn{\ds\bmv{q}(\bmv{r}_1;t)=-\frac 23n^2\sigma^4g_2
	\lp\sigma|n,\beta\rp\sqrt{\frac{\pi k^3T}m}\nabla T
	\left(\bmv{r}_1;t\right)-{}}\\
&&\lp 1+\frac 25\pi n\sigma^3g_2(\sigma|n,\beta)\rp
	\int\limits_{-\infty }^t\d t'\;\e^{-\varepsilon(t-t')}
	L^k\left(t,t'\right)\left[\frac 1T\nabla T\right]_{t'}\,.\\
\eean

Now we can consider one of the limiting cases, namely, the stationary
process, when the operator $L_t$ does not depend on time,
i.e. $S(t,t')=\exp\lc L_t(t-t')\rc$. Some terms in expressions for
$\tensor{P}(\bmv{r}_1;t)$ and $\bmv{q}(\bmv{r}_1;t)$ can acquire
simpler form. We can compare them with those from the Enskog-Landau kinetic
equation for one-component system of charged hard spheres with using the
Chapman-Enskog method in the case, when in a long-range part of the collision
integral we put $g_2\left(\sigma|n,\beta\right)\to 1$.
It should be noted that bulk viscosity has the same structure as in the
Chapman-Enskog method \cite{c8}. But other transport coefficients exhibit
some distinctions. The structure for shear viscosity $\eta$ and thermal
conductivity $\lambda$ is:
%
%
\bea
\lefteqn{\ds\eta=\frac{3}{5}\;\ae+
	2\;nkT\;\frac{\ds\lc 1+\frac{4}{15}\pi n\sigma^3
	g_2(\sigma|n,\beta)\rc^2}
	{\ds\Big\{\ds I_E^{(0)}(\delta f)+I_L(\delta f)\Big\}},}
	\label{e50}
\eea
%
%
%
%
%
\bea
\lefteqn{\ds\lambda=\frac{3k}{2m}\;\ae+
	\frac{5k}{m}\;nkT\;
	\frac{\ds\lc 1+\frac{2}{5}\pi n\sigma^3
	g_2(\sigma|n,\beta)\rc^2}
	{\ds\Big\{\ds I_E^{(0)}(\delta f)+I_L(\delta f)\Big\}}.}
	\label{e51}
\eea
Then the problem lies in calculating collision integrals
$I_E^{(0)}(\delta f)$ and $I_L(\delta f)$,
this means that we should calculate
collision integrals \refp{e2} (in the zeroth approximation on inhomogeneity)
and \refp{e4} together in the first approximation on deviation $\delta f$,
where
$\delta f$ is substituted from \refp{e31}. The matter of some difficulty is
that correction \refp{e31} in its turn is expressed also via collision
integrals $I_E^{(0)}(\delta f)$, $I_L(\delta f)$, which are in the operator
$S(t,t')$. So the first acceptable approximation should be that, when
correction
$\delta f$ \refp{e31} is expressed via $I_E^{(0)}(\delta f)$,
$I_L(\delta f)$ calculated with $\delta f'$, where $\ds\delta f'=\delta f$
at ${S(t,t')=1}$. For $I_E^{(0)}(\delta f)$ we obtain the results
\cite{c8}, for $I_L(\delta f)$ in \refp{e50}, \refp{e51} we can obtain the
following:
\bea
\lefteqn{\ds I_L(\delta f)=\frac{Z^4e^4}{m^2}
	\frac\partial{\partial\bmv{v}_1}\int\d\bmv{r}_{12}\;\d\bmv{v}_2\;
	g_2\left(\bmv{r}_1,\bmv{r}_1+\bmv{r}_{12};t\right)
	\frac{\bmv{r}_{12}\bmv{r}_{12}}{r_{12}^5}\frac 1g
	\ds\left\{\frac\partial{\partial\bmv{v}_1}-
	\frac\partial{\partial\bmv{v}_2}\right\}}\nonumber\\
\lefteqn{\ds\times\Big\{ f_1\left(x_1;t\right)
	\delta f\left(\bmv{r}_1+\bmv{r}_{12},\bmv{v}_2;t\right)+
	\delta f\left(x_1;t\right)
	f_1\left(\bmv{r}_1+\bmv{r}_{12},\bmv{v}_2;t\right)\Big\},}
	\label{e52}
\eea
where $\delta f\left(x;t\right)$ is evaluated from \refp{e31} with
\[
S(t,t')=\exp\Big\{L_t(t-t')\Big\}=
	\exp\Big\{[I_E^{(0)}(\delta f')+I_L(\delta f')](t-t')\Big\}
\]
at $\ds\delta f'\left(x;t\right)=
\delta f\left(x;t\right)\Big|_{\ds S(t,t')=1}.$
This stage of calculations needs an explicit form for the binary
quasiequilibrium correlation function $g_2$ both on the contact and in
$\bmv{r}$-space.

The results of this paper \refp{e31}, \refp{e50} and \refp{e51}
will coincide completely with those from \cite{c8} when in a long-range part
of the collision integral \refp{e52} one puts
$g_2(\bmv{r}_1,\bmv{r}_1+\bmv{r}_{12})\equiv 1$ and represents it in
Boltzmann-like form. But the used boundary conditions method has turned out
more convenient than the Chapman-Enskog one \cite{c2,c8}. As was discussed
in details in \cite{c12} at constructing the normal solution for a kinetic
equation using the boundary conditions method, time derivatives
$\partial/\partial t$ of hydrodynamic parameters of reduced description do
not set to be small. Therefore, the normal solution to this equation could
be used for hydrodynamic description of fast processes.

\end{document}